\documentclass[amsmath,twocolumn,groupaddress,prb,aps, amsmath,amssymb,amsfonts,longtable]{revtex4-1}
\usepackage{amsthm,amsfonts,graphicx,verbatim, color}
\usepackage{bm}
\usepackage{amsmath}
\usepackage{graphicx}
\usepackage{braket}
\usepackage{dcolumn}
\usepackage{bbm}
\usepackage{subfigure}
\usepackage{amssymb}
\usepackage{hyperref}
\usepackage{wasysym}
\usepackage{color}
\usepackage{MnSymbol}
\usepackage{tabularx}
\stepcounter{secnumdepth}
\stepcounter{tocdepth}
\usepackage[utf8]{inputenc}
\usepackage{ulem}

\newcommand{\mc}{\mathcal}

\begin{document}

\title{AC Conductivity Crossover in Localized Superconductors}
\author{A.T. Schmitz}
\affiliation{Department of Physics and Center for Theory of Quantum Matter, University of Colorado, Boulder, CO 80309, USA}
\author{Michael Pretko}
\affiliation{Department of Physics and Center for Theory of Quantum Matter, University of Colorado, Boulder, CO 80309, USA}
\author{Rahul M. Nandkishore}
\affiliation{Department of Physics and Center for Theory of Quantum Matter, University of Colorado, Boulder, CO 80309, USA}

\begin{abstract}
An important experimental signature of localization is the low-frequency AC conductivity, which typically vanishes as $\omega^{\phi}$. The exponent $\phi = 2$ for Anderson insulators, whereas for many body localized insulators $\phi$ is a continuously varying exponent $1 \le \phi \le 2$.  In this work, we study the low-frequency AC conductivity of localized {\it superconductors}, in which disorder is strong enough to localize all quasiparticles, while remaining weak enough to leave superconductivity intact.  We find that while the ac conductivity still follows the general form $\sigma(\omega) \sim \omega^{\phi}$, the exponent $\phi$ can be markedly different from the characteristic value for localized insulators.  This difference occurs due to singularities in the low-energy density of states, permitted by the effective particle-hole symmetry around the Fermi level.  In particular, in certain symmetry classes at zero temperature, we obtain $\phi > 2$. We further identify an interesting temperature dependent crossover in the scaling form of the AC conductivity, which could be useful for the experimental characterization of localized superconductors.
\end{abstract} 
 \maketitle

\tableofcontents

\section{Introduction}

Many-body localization (MBL) \cite{pwa, fa, agkl, gornyi,basko1,imbrie,ogan,xxz, pal, review} has become a subject of intense theoretical\cite{moore,bauer,serbyn,swingle,order,growth,spt,huse,marginal,rahul,laumann,meanfield,tarun,khemani,protection,ros,gopal,bahri,vosk,finite,criterion,ngh,randomfield,universal,proximity,symmetry,deroeck,spectral,zhang,critical,chandran,bath,spectrum,power,italian,stability,highly,continuum,gaugembl,zhicheng,dumitrescu,class,geraedts,abhinav,spectrum2,acevedo,mobile,longrange,mottglass} and experimental\cite{exp1,exp2,exp3,exp4, schneider, monroe, lukin, capellaro, rosenbaum, grant, mihailovic} interest in recent years as an example of a mechanism by which an interacting many-body system can violate the ergodic hypothesis. In localized phases, the system retains a memory of its initial conditions in local observables at infinitely long times, thereby failing to reach thermal equilibrium.  While localized phases are usually thought of as insulators, localization is in principle also compatible with superconductivity - after all, superconductivity is a property of the {\it ground} state, and localization is a property of the excitations. Discussions of `Anderson localized' superconductors (in which the Bogolioubov-de-Gennes quasiparticles are localized) date back at least as far as Refs.\onlinecite{Senthil1999,Senthil2000,Vishveshwara2000,Vishveshwara2001}. More recently, it has been argued \cite{longrange, extended} that MBL is also compatible with superconductivity, and indeed could be used to stabilize superconductivity to energy densities where in equilibrium the phenomenon would not arise.

How could a localized superconductor be experimentally characterized? One possibility \cite{Senthil1999,Senthil2000,Vishveshwara2000,Vishveshwara2001} is through dc transport: the system should be a superconductor for charge, but a thermal insulator. However, a dc measurement typically requires that contacts be attached, whereas in the modern theory of localization one ideally wants to treat the system as a closed quantum system. AC conductivity offers an alternative route to experimental characterization, and has the advantage of being accessible in purely optical (contact free) measurements. Localized insulators have well understood signatures in AC conductivity. For non-interacting Anderson insulators, the AC conductivity vanishes at low frequency according to Mott's law \cite{Mott1968}, $\sigma \sim \omega^2$, with logarithmic corrections that we will not discuss here. For many body localized systems, the conductivity still vanishes \cite{gopal} as $\sigma \sim \omega^{\phi}$, but with $\phi$ a continuously varying exponent $1 \le \phi \le 2$, which approaches $2$ deep in the localized phase, and approaches $1$ close to the delocalization transition. Indeed the possibility of interaction driven modifications to Mott's law was noted already in the classic work Ref.\onlinecite{ES} in the context of systems with Coulomb interactions. However, the AC conductivity of a localized {\it superconductor} has yet to be worked out.

In this paper, we study the AC conductivity of localized superconductors. We focus throughout on the real part of the conductivity, and our results are accurate up to logarithmic corrections, as in Ref. \onlinecite{gopal}.  We begin by discussing the low-frequency AC conductivity of a superconductor in the Anderson-localized regime of the Bogoliubov quasiparticles. For certain symmetry classes we find a modification to Mott's law at zero temperature. Specifically, we find that the conductivity vanishes as $\sigma \sim \omega^{\phi}$ with $\phi > 2$.  This modification is a result of singularities in the low-energy density of states, which are allowed by the effective particle-hole symmetry around the Fermi level.  We argue that at zero temperature, the Anderson-localized result should be unaffected by turning on interactions between quasiparticles. We then consider the effects of non-zero energy density (given non-vanishing interactions between quasiparticles), adapting the arguments of Ref. \onlinecite{gopal}. We find a temperature driven crossover in the scaling behavior controlled by a parameter $\alpha^{-1}$ which can be interpreted as the thermal average number of resonant localized Bogoliubov quasiparticle states within a localization volume. The crossover is sharper in higher dimensions. We then consider the contribution of vortices and plasmons. We argue that neither type of excitation contributes to the bulk AC conductivity in linear response. Our analysis of the quasiparticle sector thus fully captures the AC linear response conductivity of localized superconductors, and provides a means of characterizing this phase, and distinguishing it from conventional localized insulators.

\section{Mott Argument for AC Conductivity of Anderson localized superconductors}

The AC conductivity of a disordered, Anderson-localized electronic system near zero frequency was first investigated by Mott\cite{Mott1968} and found to be $\sigma(\omega)\sim \omega^2$, up to logarithmic corrections. As our argument for a dirty BCS superconductor closely follows Mott's, we review the argument in detail.

We start with the single-particle Kubo formula for AC conductivity in a non-interacting system, also known as the Kubo-Greenwood formula
\begin{align}
\sigma_{s.p.}(\omega,T) =& \frac{\pi}{V\omega} \sum_{\alpha, \beta} \braket{\alpha|\vec{j}|\beta}\cdot \braket{\beta|\vec{j}|\alpha} \delta\left(\omega- (E_\alpha -E_\beta)\right)\nonumber \\
&\times \left(f(E_\alpha, T)- f(E_\beta, T)\right),
\end{align}
where $\ket{\alpha},\ket{\beta}$ are the single particle eigenstates of the system with energy $E_\alpha, E_\beta$ respectively, $\vec{j}$ is the total current operator (sum over local current operators), $V$ is the system volume, $f(E,T)$ is the Fermi-Dirac distribution function, and the subscript $s.p.$ denotes ``single-particle." In a semi-classical approximation, $\vec{j} \sim \frac{d\vec{x}}{dt} \sim [H, \vec{x}]$, where $H$ is the single particle Hamiltonian for the system. Thus the current matrix elements should go as 
\begin{align}\label{eq:jelm}
|\braket{\alpha|\vec{j}|\beta}| \sim |\braket{\alpha|[H,\vec{x}]|\beta}| \sim |E_\alpha - E_\beta| |\braket{\alpha|\vec{x}|\beta}|.
\end{align}
For a localized system, $\ket{\alpha}$ and $\ket{\beta}$ are centered at some position with an exponentially decaying wave function in real space. Furthermore, two states that are close together in real space must have a large energy difference. For if they did not, the near-resonance would cause hybridization of their wave functions and this would destroy localization. Thus for the energy difference between $\ket{\alpha}$ and $\ket{\beta}$ to be $\omega$, the localization centers of these states should be separated by a distance $r_\omega$, given by $ \omega \sim W \exp\left(-\frac{r_{\omega}}{\xi}\right)$, where $\xi$ is the localization length and $W$ is the bandwidth. Thus we can make the replacement $|\braket{\alpha|\vec{x}|\beta}| \sim r_\omega \sim -\ln(\omega)$. Plugging these results into the Kubo formula and assuming the matrix elements are roughly constant over all relevant states (those with small energy differences), we obtain

\begin{align}
\sigma(\omega, T) \sim\,& \omega r_{\omega}^2 r_{\omega}^{d-1} \sum_{\alpha, \beta}\delta\left(\omega- (E_\alpha -E_\beta)\right) \nonumber \\
&\times \left(f(E_\alpha, T) - f(E_\beta, T)\right) \nonumber \\
\sim\,&\omega\ln^{d+1} (\omega)\int dE_\alpha dE_\beta \rho(E_\alpha)\rho(E_\beta) \nonumber \\
&\times \left(f(E_\alpha,T) - f(E_\beta,T)\right) \delta(\omega - (E_\alpha - E_\beta)) \nonumber \\
\sim\,& \omega \ln^{d+1} (\omega) \int dE \,\rho(E + \omega) \rho(E) \nonumber \\
&\times \left(f(E + \omega, T) - f(E, T)\right),
\end{align}
where $\rho(E)$ is the single particle density of states, and the factor of $r_{\omega}^{d-1}$ comes from the number of possible transitions contributing to the conductivity.  Note the presence of two factors of the density of states at two different energies.  This is to be expected, since a transition between two energy levels naturally involves the density of states at both the final and initial energies.  Also note that we have taken the localization length to be approximately constant at low energies, as argued in Reference \onlinecite{Senthil1999}.

We can now use this expression to obtain the standard Mott's law for AC conductivity.  There are two separate low-frequency limits, both of which we will consider.  First, let us take the temperature to zero first, then take $\omega \rightarrow 0$.  For temperatures near zero, the Fermi-Dirac function looks like an inverted step function at the Fermi energy $\mu$ which for simplicity we take to be zero. Thus the combination $ \left(f(E + \omega, T) - f(E, T)\right)$ is roughly a rectangle function of width $\omega$, so that the zero-temperature AC conductivity takes the form

\begin{align} \label{eq:kuboredux}
\sigma(\omega, T \approx 0) \sim \omega \ln^2 (\omega) \int_0^\omega dE \rho(E + \omega) \rho(E).
\end{align}
Typically, we expect the leading behavior of the DOS near the Fermi energy to be a constant in which case we recover $\sigma \sim \omega^2$ (dropping the inconsequential log factors). This is Mott's law.  One can also find similar behavior in the limit of fixed temperature, taking $\omega\rightarrow 0$ first.  In this $\omega\ll T$ limit, the Fermi function difference goes as $\omega/T$ in a first order Taylor expansion, while the energy window over which it is non-negligible is now controlled by $T$ instead of $\omega$.  In this case, we can once again conclude that $\sigma(\omega\rightarrow 0,T) \sim \omega^2$, just as at zero temperature.

Now let us generalize this argument to superconductors.  We first consider the low-temperature limit, with a generalization to the high-temperature limit discussed later.  To analyze the conductivity of a superconductor, we must first write the current operator in terms of Bogoliubov quasiparticles, the natural excitations of the superconductor. In terms of the electron operators $c_{\vec{k} \uparrow},c_{\vec{k} \downarrow}$, the current operator is 
\begin{align}
\vec{j} = \sum_{\vec{k}, \sigma} \vec j_{\vec{k}} c^\dagger_{\vec{k} \sigma} c_{\vec{k} \sigma}, \label{barecurrent}
\end{align}
where $\vec{j}_{\vec{k}} \sim \vec{k}$ at low momentum, but regardless of the precise form is an odd function of $ \vec{k}$. We can then rewrite the current in Bogoliubov quasiparticle operators, $\gamma_{\vec{k} 0}, \gamma_{\vec{k}1}$, which satisfy the canonical transformation relations \cite{Tinkham2004}
\begin{subequations}\label{eq:BdG}
\begin{align}
c_{\vec{k}  \uparrow} = u_{\vec{k}} \gamma_{\vec{k} \uparrow} + v_{\vec{k}} \gamma_{-\vec{k} \downarrow}^\dagger,\\
c^\dagger_{-\vec{k} \downarrow}= -v_{\vec{k}}^* \gamma_{\vec{k} \uparrow} + u_{\vec{k}} \gamma_{-\vec{k} \downarrow}^\dagger,
\end{align}
\end{subequations}
where $u_{\vec{k}}, v_{\vec{k}}$ are the usual quasiparticle transformation coefficients which must satisfy $|u_{\vec{k}}|^2 + |v_{\vec{k}}|^2 =1$ for the quasiparticles to be normalized fermions. Using Eqs. \eqref{eq:BdG}, the normalization relation, the Bogoliubov quasiparticle anti-commutation relations, and the anti-symmetry of $\vec{j}_{\vec{k}}$, it is easy to show that
\begin{equation}
\vec{j} = \sum_{\vec{k}, \sigma} j_{\vec{k}} \gamma^\dagger_{\vec{k} \sigma}\gamma_{\vec{k} \sigma} 
\end{equation}
so the current operator takes exactly the same form when written in terms of Bogolioubov quasiparticles as when written in terms of electrons. This is somewhat surprising as Bogoliubov quasiparticles do not themselves carry charge, but is a known result in the literature (see Ref. \onlinecite{Nayak2001}). Given this observation, we can simply repeat Mott's arguments to conclude that, when the Bogolioubov quasiparticles are Anderson localized, then the AC conductivity must be given by Eq. \ref{eq:kuboredux}. {\it However}, unlike the case of free electrons, the density of states for Bogolioubov quasiparticles need not be a constant at low energies. We now adapt this reasoning to superconductors. Suppose the DOS goes as some power law near the Fermi-surface with $ \rho(E) \sim E^\eta (\eta \geq 0$). Precisely this kind of dependence will arise in certain superconductors. It is then clear Eq. \eqref{eq:kuboredux} that
\begin{align} 
\sigma(\omega) \sim \omega^{2 \eta +2}, 
\end{align}
 for $T=0$ where we  have dropped the log factors. This is sharply distinct from Mott's law, and also from its many-body generalization \cite{gopal}, in that it constitutes a scaling form $\sigma \sim \omega^{\phi}$ with $\phi>2$. 

Another possibility is a log divergent DOS near the Fermi surface with the form $\rho(E)\sim \left(-\ln (E)\right)^\beta$. To approximate the DOS integral, we can write the following log factor as
\begin{align}
 \ln(E + \omega) =\left( \ln \left(1 + \frac{E}{\omega}\right) + \ln (\omega)\right).
\end{align}
 Here, the $\ln(\omega)$ term is going to dominate over the  $\ln \left(1 + \frac{E}{\omega}\right)$ term as this only ranges over the limits of integration from zero to $\ln(2)$. In that case, one can make the approximation
\begin{align}
\ln(E) \ln(E + \omega)\approx  \ln(\omega) \ln(E).
\end{align}
So the integral can be approximated by
\begin{align} \label{eq:logint}
&\int_0^\omega dE  \left(-\ln(E)\right)^\beta  \left(- \ln(E + \omega)\right)^\beta\nonumber \\
&\approx \left(- \ln(\omega)\right)^\beta \int_0^\omega\left(-\ln(E)\right)^\beta  dE \nonumber \\
&= (-\ln(\omega))^\alpha \int_{-\ln(\omega)}^\infty \exp(-u)u^\alpha du \nonumber \\
 &\sim \omega (-\ln(\omega))^{2\beta}. 
\end{align}
Including the other factors from the Kubo formula in Eq. \eqref{eq:kuboredux}, we recover the Mott law with additional log corrections.

\section{AC conductivity of superconductors with Anderson localized quasiparticles} 

The dirty superconductor in the BCS regime is a well-studied model in the literature. While certain symmetry classes have constant density of states in the low energy limit, others have power law or logarithmic behavior that will manifest in AC conductivity. Here we take up the symmetry classes discussed in Refs. \onlinecite{Senthil1999,Senthil2000,Vishveshwara2000,Vishveshwara2001}. In Refs. \onlinecite{Senthil1999,Vishveshwara2000,Vishveshwara2001}, the authors discuss a phase transition within a dirty superconductor at low temperature and varying magnetic field.  In particular, they study three-dimensional models with an approximate $SU(2)$ symmetry, approximate only in that the Zeeman term is negligible.  (It is argued that there is no analogous transition within two-dimensional superconductors in this symmetry class.)  At the highest field strengths, one has a normal metal. As the field decreases, the system first goes into a ``thermal metal'' phase which is superconducting as well as thermally conducting due to delocalized Bogoliubov quasiparticles.  Upon further reducing the field, the system eventually enters the ``thermal insulator'' phase where the quasiparticles are also localized and one has no diffusive transport.  For the thermal metal, one finds $\rho(E) \sim \rho_0 + \sqrt{E}$. If $\rho_0$ is significantly greater that zero, than one just get Mott's law. If this is not the case, then one will obtain $\sigma(\omega) \sim \omega^3$. However, one generally expects $\rho_0 \neq 0$. 

For the thermal insulator phase, one can start modeling the system by considering the case of infinite onsite disorder. This results in a decoupled Hamiltonian composed of onsite terms $h_{\vec{x}} = \vec{a}_{\vec{x}} \cdot \sigma$, where $\vec{a}_{\vec{x}}$ is a random vector.  If $\vec{a}_{\vec{x}}$ can point in any direction, then the constant energy density surface is represented by a sphere in $\vec{a}_{\vec{x}}$ space and one has that $\rho(E) \sim E^2$. However, if one has time reversal symmetry, $\vec{a}_{\vec{x}} \cdot \hat{y}=0$ and the constant energy surface becomes a circle, implying $\rho(E) \sim E$. Numerical simulations presented in those papers confirm this argument. These behaviors for the DOS give $\sigma(\omega) \sim \omega^4$ or $\omega^6$ for the cases with and without time reversal symmetry, respectively.

As another important case, in Ref. \onlinecite{Senthil2000}, the authors discuss two-dimensional systems with the addition of spin-orbit coupling in which case we have no spin symmetry.  The spin-orbit coupling allows for a thermal metal-insulator transition in two dimensions. The insulator case with time-reversal symmetry follows a similar argument as before where we find that $ \rho \sim E$. However when time reversal symmetry is broken the DOS goes as a constant. The more dramatic difference is in the thermal metal phase (here low disorder). They show via replica arguments that

\begin{align}
\rho(E) \sim \begin{cases}
\sqrt{-\ln(E)},& \text{ time-reversal}\\
-\ln(E), & \text{no time-reversal}
\end{cases}.
\end{align}
We can then apply Eq. \eqref{eq:logint} to find the Mott result with extra log corrections.
 (The authors also briefly discuss three-dimensional systems, conjecturing that the density of states goes as a constant.)

\begin{table*}[t]
\begin{ruledtabular}
\begin{tabular}{c c c l l}
Dimension & Symmetries &Phase&  $\rho(E)$ & $\sigma(\omega)$\\
\hline
$3$ &$SU(2), \mc T$ & Thermal Insulator (localized) &$\sim E$ & $\sim \omega^4$ \\
$3$ & $SU(2),$ (no $ \mc T$) &Thermal Insulator (localized) & $\sim E^2$ & $\sim \omega^6$ \\
$2,3$ & $U(1), \mc T$ or no $\mc T$ &Thermal Insulator (localized) & $\sim const$ & $\sim \omega^2$ \\
$2$ & No spin sym, $\mc T$ &Thermal Metal (delocalized) & $ \sim \sqrt{-\ln(E)}$ & $ \sim \omega^2$ (ELC) \\
$2$ & No spin sym, $\mc T$ & Thermal Insulator (localized) &$ \sim E$ & $ \sim \omega^4$ \\
$2$  &No spin sym,  no $\mc T$&Thermal Metal (delocalized) & $ \sim -\ln(E)$ & $ \sim \omega^2$ (ELC)\\
$2$  &No spin sym,  no $\mc T$& Thermal Insulator (localized) &$ \sim const$ & $ \sim \omega^2$
\end{tabular}
\caption{Summary of non-interacting results at $T=0$. $\mc T$ is time-reversal symmetry. ELC stands for ``extra log correction'', i.e. those not given by the constant DOS Mott's law.} \label{table}
\end{ruledtabular}
\end{table*}

An important case not in the literature is the case of spin $U(1)$ symmetry. However, the above argument for the $SU(2)$ symmetric case can be adapted here. In the infinite onsite disorder limit, the random $\vec{a}_{\vec{x}}$ for the site at $\vec{x}$ must point in some specific direction. This limits the possible allowed values of $\vec{a}_{\vec{x}}$ to a line, where the constant energy density surface is two points. Thus this has no energy dependent contribution to the DOS, and we expect $\rho(E) \sim const$. This yields the Mott result once again. Table \ref{table} summarizes these results as well as the conductivity behavior they imply. We note that some cases which differ from the Mott law in an appreciable way are the three-dimensional systems with $SU(2)$ spin symmetry. 

According to Ref. \onlinecite{symmetry}, systems with nonabelian symmetries, such as spin $SU(2)$ symmetry, do not have a stable many body localized phase. However, these arguments do not apply to the {\it Anderson} localized scenario that we are considering at present. Additionally, the $SU(2)$ symmetry being considered here is only approximate, and below a small energy scale set by Zeeman coupling strength, we expect a crossover to $U(1)$ behavior, such that there is no true non-Abelian symmetry in the problem. Finally, even if the assume the $SU(2)$ symmetry is perfect, and that interactions between quasiparticles are present, the arguments in \onlinecite{symmetry} really only rule out a stable localized phase at {\it infinite} temperature. Generally, we would expect superconducting systems to have a mobility edge where high energy states are thermal (and not superconducting). We are not aware of any obstructions to having localization in the low energy Hilbert space of a model with non-Abelian symmetry. All arguments in this paper are modulo rare regions. We defer discussion of rare region effects, and whether they obstruct MBL mobility edges and/or MBL in dimensions greater than one, to future work. Even if rare regions do pose such obstructions, they would only become apparent on extremely long lengthscales and timescales, and our results should remain accurate up to such scales.

\section{AC Conductivity in Superconducting MBL Systems}
We now turn on short-range interactions between quasiparticles, and ask how these alter our results. We continue to neglect vortices and plasmons, which (as we will later show) do not contribute to AC conductivity in linear response. 

\subsection{Zero temperature}
At zero temperature, the quasiparticles are present with vanishing density. Each quasiparticle may thus be treated as effectively non-interacting, and our previous results for superconductors with Anderson localized quasiparticles continue to apply. 

\subsection{Infinite-Temperature}

We now consider the infinite temperature limit (assuming for now that superconductivity survives to infinite temperature). Here we must use the full many-body Kubo formula
\begin{align}\label{eq:MBLkubo}
\sigma_{m.b.}(\omega,T) =& \frac{1- \exp \left(-\frac{\omega}{T}\right)}{\omega Z} \sum_{\alpha, \beta} \braket{\alpha|\vec{j}|\beta}\cdot \braket{\beta|\vec{j}|\alpha}\nonumber \\
& \times \delta\left(\omega- (E_\alpha -E_\beta)\right) \exp\left(-\frac{E_\alpha}{T}\right),
\end{align}
where $\alpha, \beta$ index the many-body energy eigenstates, $Z$ is the partition function, and the subscript $m.b.$ denotes ``many-body." Taylor expansion reveals that the conductivity vanishes in the infinite temperature limit as $1/T$, so we calculate instead $T \sigma(\omega)$. The current operator in this expression no longer hops a single localized Bogoliubov quasiparticle as before but rather connects states differing by entire configurations of localized Bogoliubov quasiparticles. Borrowing from arguments made in Ref. \onlinecite{gopal}, this has the effect of enhancing the density of resonant states at infinite temperature by an exponential factor,
\begin{align} \label{eq:MBDOS}
\rho_{MB} (E+\omega) \rho_{MB} (E) \sim \rho (E+\omega) \rho(E) \exp(sn),
\end{align}
where $n$ is the number of Bogoliubov single particle states which have changed between resonant many-body states (what we shall refer to as ``flips") dominating the sum in Eq. \ref{eq:MBLkubo}.  The quantity $s$ is an entropy density for these dominant resonances. The number of flips in a dominant resonance follows from an energy difference argument analogous to Mott 
but for resonant flips and is given by
\begin{align} \label{eq:MBom}
\omega \sim W \exp\left(-\frac{n}{\zeta}\right),
\end{align}
where $\zeta$ is a unit-less parameter analogous to the localization length and is determined by the level of disorder in the Hamiltonian. We expect the current-operator matrix elements still go as $\omega$ but in the infinite temperature limit, any effects due to the single-particle DOS near the Fermi surface are washed out, leaving
\begin{align}
T \sigma (\omega) \sim \omega^{2 - \varphi},
\end{align}
where $\varphi =s \zeta \in [0,1]$ so that $\phi= 2- \varphi$. This result is indistinguishable from the infinite temperature AC conductivity of MBL electronic systems, discussed in Ref. \onlinecite{gopal}. Of course, we do not expect superconductivity to persist to infinite temperature anyway. 

\subsection{Finite non-zero temperature}
\begin{table}[t]
\begin{tabular}{c | c}
Variable & Definition\\[.25em]
\hline \\[.1em]
$\alpha$ & $(\xi^d \varrho)^{-1}$\\[.5em]
$\Omega$ & $\ln \left(\frac{W}{\omega}\right)$\\[.5em]
$\Omega_c$ & $ \left(\zeta \alpha\right)^{\frac{1}{d-1}}$\\[.5em]
$\omega_c$ & See Eq. 22\\[.5em]
$\overline \Omega$ & $\frac{\Omega}{\Omega_c}$\\[.5em]
$\overline \Omega_\epsilon $ & see Eq. \protect \eqref{eq:eps}
\end{tabular}
\caption{List of variables used in this section. Here $\xi$ is the localization length and $\varrho$ is the density of flips. } \label{table2}
\end{table}
We now consider finite non-zero temperatures (assumed to be low enough for the system to still be superconducting). In this case there are two regimes, with a crossover between the two. (An analogous crossover in the context of systems with a Coulomb interaction was discussed in Ref.\onlinecite{ES}).  At first glance, one might think that frequencies small compared to temperature would follow the `infinite temperature' scaling, and frequencies large compared to temperature would follow the `zero temperature' scaling. However, temperature also controls the thermodynamic weight of the terms in Eq. \eqref{eq:MBLkubo}. Moreover, these weights depend on the absolute energy of the states as it compares to temperature, instead of just the relative energy difference between the two states. So a finite temperature, even if large compared to the the frequency, can still have an effect on the general behavior of AC conductivity. 

To understand the effects of finite temperature, we revisit the energy difference relation given in Eq. \eqref{eq:MBom}. This equation assumes energy difference only depends on the number of flips, but generally we also expect the energy to depend on the distance between those flips. So in terms of the absolute energy of the final and initial states, configurations of flips which occur over larger volumes should have an exponentially smaller absolute energy as compared to those which occupy less volume. To reiterate, this does not matter for the infinite temperature case because the absolute energy of the initial and final states is irrelevant due to even statistical weighting in the Kubo formula. The radius defining the volume of the flips should scale as $n^{\frac{1}{d}}$ in $d$ spatial dimensions, in which case we have an effective energy relation
\begin{align}\label{eq:MBom2}
\omega \sim W \exp\left(-\frac{n}{\zeta}\right) \exp\left(-(\alpha n)^{\frac{1}{d}}\right),
\end{align}
where $\alpha$ is a statistical parameter which is dependent on temperature and the localization length and characterizes the average radius of the smallest volume which captures all $n$ flips between dominant resonant states. To understand how $\alpha$ depends on temperature and localization length, consider when we can capture all flips between two resonant energy eigenstates within a ball of radius $r$. Then we expect $(\alpha n)^{\frac{1}{d}} \sim \frac{r}{\xi}$. From this we can say that
\begin{align}
\alpha^{-1} \sim \xi^d \left(\frac{n}{r^d}\right) \sim \xi^d \varrho
\end{align}
where $\varrho$ is the density of flips. So $\alpha^{-1}$ is roughly the average number of flips within a localization volume. This can also be thought of as the effective interaction strength.  When there are several flips within a localization volume, the interactions between those states becomes more relevant, whereas when there are one or less flips within a localization volume, the behavior should be more like Anderson localization.

To estimate the dependence of $\varrho$ on temperature, we can count the number of states with energy less than the temperature using the DOS, $\varrho \sim \frac{1}{V} \int_0^T dE \rho(E) \sim\frac{T^{\eta +1}}{V}$, for a system of volume $V$ and low temperature. So the low-temperature behavior of $\alpha$ is
\begin{align}
\alpha\sim \frac{V}{\xi^d} T^{-(\eta+1)}.
\end{align}
For general behavior with respect to temperature, $\alpha$ should be smooth and monotonically decreasing with increasing temperature. Thus $\alpha^{-1}$ will be monotonically {\it increasing} with temperature. At low temperature, $\alpha^{-1} \ll 1$ (effectively Anderson localized behavior), and at high temperatures $\alpha^{-1} \gg 1$ (behavior as in Ref. \onlinecite{gopal}). 

Using Eq. \eqref{eq:MBom2}, we can now write an equation for $n$
\begin{align}\label{eq:neq}
\left(\Omega -\frac{n}{\zeta} \right)^d - \alpha n \sim 0,
\end{align}
where $\Omega = \ln\left(\frac{W}{\omega}\right)$. A general real solution for $n$  in any $d\geq 2$ is
\begin{align}\label{eq:n2}
n \sim& \zeta\left(\Omega- \Omega_c \left(F_d(\overline \Omega)\right)^\frac{1}{d}\right)  \nonumber \\
=& \zeta\left(\Omega - \Omega_c \left( \overline \Omega \right)^{\frac{1}{d}}\right) + \mc O\left( \overline \Omega^{\frac{2-d}{d}}\right).
\end{align}
where $\Omega_c^{d-1}= \zeta \alpha$, $\overline \Omega = \frac{\Omega}{\Omega_c}$ (as also expressed in Table \ref{table2}) and $F_d(x)$ is the real solution to the polynomial equation $(x- F_d)^d -F_d =0$ such that $F_d(0) =0$, which always exists. One can see that $F_d(x) \to x$ as $x \to \infty$ so the last expression of Eq. \eqref{eq:n2} is valid in the limit that $\Omega \gg \Omega_c$. (Also note $F_d$ is easily invertible, namely $F_d^{-1}(x) = x + x^\frac{1}{d}$.) From $\Omega_c$, we can also extract a frequency
\begin{align}
\omega_c = W \exp\left(- (\zeta \alpha)^{\frac{1}{d-1}}\right)= W\exp \left(- \left(\frac{r_{\text{eff}}}{\xi}\right)^{\frac{d}{d-1}}\right),
\end{align}
where $r_{\text{eff}}$ is some temperature-dependent effective distance. So the deeper in the MBL phase, the smaller the critical frequency.

We can now use Eq. \eqref{eq:n2} in the exponential factor for resonant density of states,
\begin{align}
\exp(sn) \sim& \exp\left( \varphi\left(\Omega- \Omega_c \left(F_d(\overline \Omega)\right)^\frac{1}{d}\right)\right) \nonumber \\
=& \left(\frac{\omega}{W}\right)^{- \varphi \left( 1-  \frac{\left(F_d(\overline \Omega)\right)^\frac{1}{d}}{\overline \Omega}\right)}
\end{align}
%
%
As we are at small but still finite temperature, we can combine the result with the Mott factors to get
\begin{align}\label{eq:COsig}
T \sigma(\omega, T) \sim
 \omega^{2- \varphi \left( 1-  \frac{\left(F_d(\overline \Omega)\right)^\frac{1}{d}}{\overline \Omega}\right)}
\end{align}

\begin{figure}[t]
\centering
\includegraphics[scale=.61]{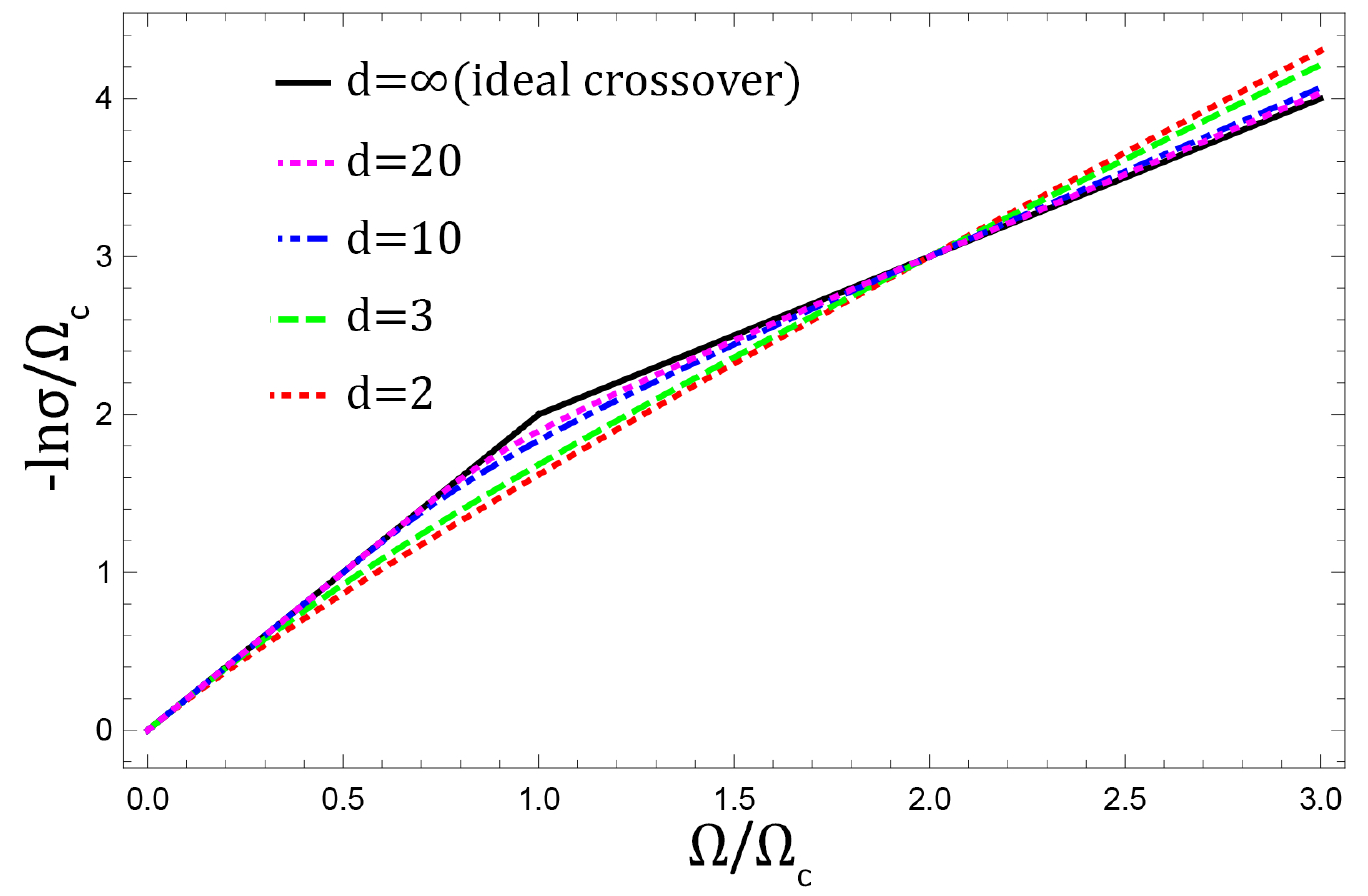}
\caption{Log-log plot of the conductivity comparing finite-temperature results in different dimensions. In all plots  $\varphi= 1$ to emphasize the crossover behavior. $d=10, 20$ is shown to demonstrate that the crossover behavior becomes ideal as $d\to \infty$. Note that all solutions cross at a single point due to the fact that $F_d(2) =1$ for all $d$.} \label{cond}
\end{figure}

Thus we find the general low-frequency behavior as stated in Ref. \onlinecite{gopal} but with a crossover from a distinctly different form as dictated by $\Omega_c$ and by extension $\alpha$.  To see the crossover behavior more clearly, Consider the log-log plot of the expected behavior.
If we then measure this in units of $\Omega_c$, we find the relation
%
%
\begin{align}
\frac{- \ln \sigma}{\Omega_c} \sim g(\overline \Omega) = (2 -\varphi) \overline \Omega + \varphi \left(F_d(\overline \Omega)\right)^\frac{1}{d}.
\end{align}
A plot of this function is shown in Fig. \ref{cond}. There we can see that at low values of $\overline \Omega$ (in which case $\frac{\omega}{W} \sim \left(\frac{\omega_c}{W}\right)^{\overline \Omega}$, so $\omega \gg \omega_c$), the behavior is that of an Anderson-localized superconductor. As $\overline \Omega$ is increased ($\omega$ is lowered relative to $\omega_c$), the behavior crosses over to the infinite temperature behavior where the slope is reduced by $\varphi$. The point of the crossover is roughly $\Omega_c$ which becomes more pronounced in higher dimensions and approaches an ideal crossover as $d\to \infty$. As the crossover point is determined by $\Omega_c$ and by extension $\alpha \sim T^{-\eta+1}$, we can see that the frequency at which this crossover behavior is observed is pushed to lower frequency as the temperature decreases. 

\begin{figure}[t]
\centering
\includegraphics[scale=.61]{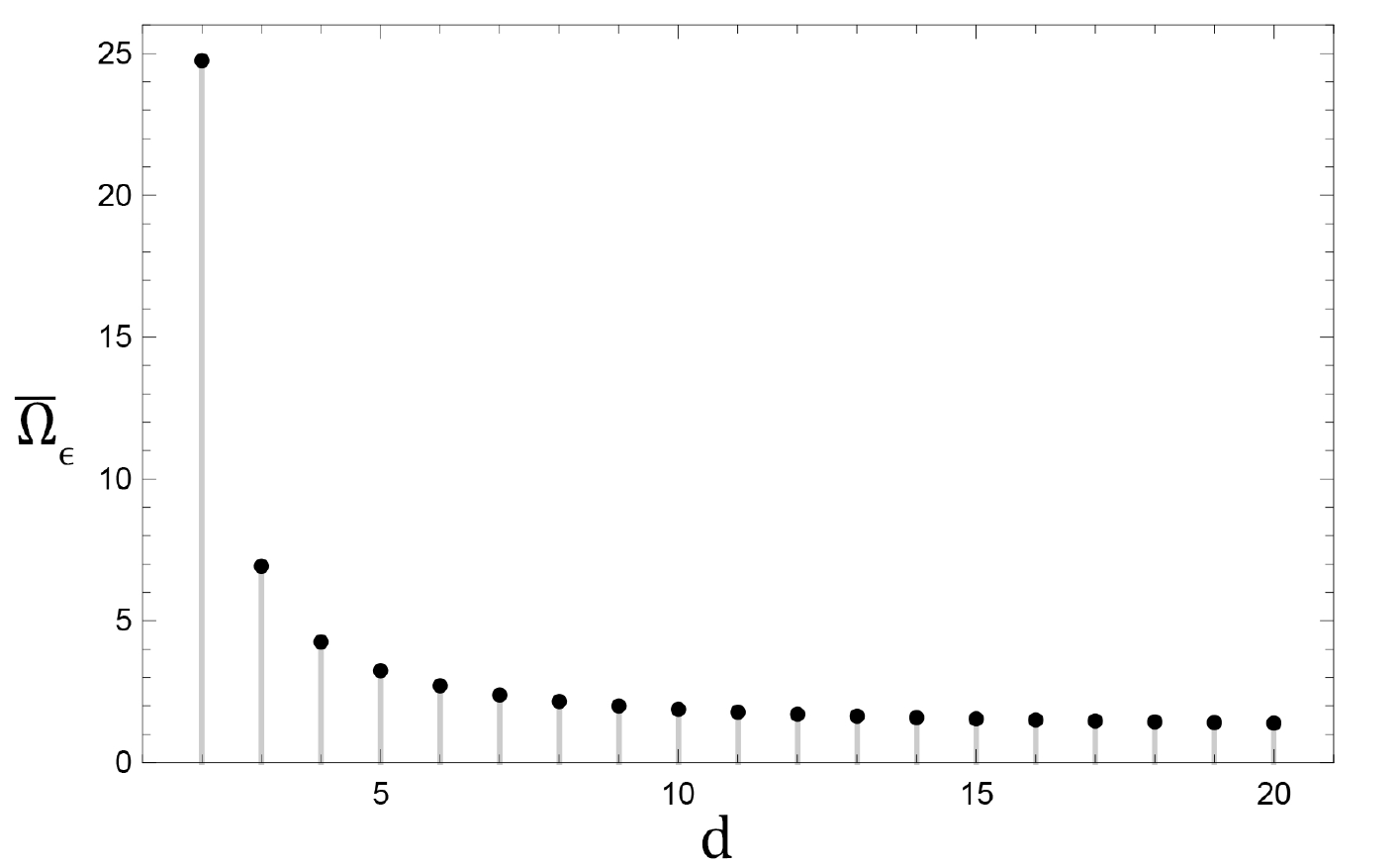}
\caption{Plot of $\overline \Omega_\epsilon$ for $\epsilon= 0.1$. Here again $\varphi= 1$.} \label{epplot}
\end{figure}

From Fig. \ref{cond}, we can see that the crossover is much softer for lower dimensions. Eventually the behavior becomes infinite temperature like, but we can determine how large $\overline \Omega$ must be (or alternatively how small $\omega$ must be) before one sees the infinite temperature like behavior. Suppose we wish to know the value $\overline \Omega_\epsilon$ for which the slope of $g(\overline \Omega)$ is within $\epsilon$ of its value at infinity, i.e.
\begin{align} \label{eq:eps}
\frac{g'(\overline \Omega_\epsilon) - g'(\infty)}{g'(\infty)} = \epsilon,
\end{align}
One can use the defining equation of $F_d$ and its inverse to find that in $d$ dimensions
\begin{align}
\overline \Omega_\epsilon = F_d^{-1} \left(\left(\frac{\varphi}{\epsilon g'(\infty) d}- \frac{1}{d}\right)^{\frac{d}{d-1}}\right).
\end{align}
Fig. \ref{epplot} plots this value for various dimensions. So for $d=2$, $\Omega \approx 25 \Omega_c$ before the measured slope is within $10 \%$ of its value at infinity.

For $d=1$, we note that superconductivity is generally not stable\cite{Tinkham2004}. Moreover even if it were, the solution to Eq. \eqref{eq:neq} in $d=1$ is easily solved in exact form as 
\begin{align}
n \sim  \frac{\zeta\Omega}{1 + \zeta \alpha}.
\end{align}
This yields the same form for AC conductivity as the infinite temperature solution with at most a modification to $\varphi$ which is still within the range $[0,1]$. This is consistent with the results of Ref. \onlinecite{Parameswaran2016}. So in one dimension, the crossover behavior does not occur, and the behavior is generically `infinite temperature like.' In dimensions larger than one, however, there is a regime of frequencies large compared to temperature in which the response is effectively that of a superconductor with Anderson localized quasiparticles. 

\section{Discussion and conclusions}

The proceeding arguments only account for one type of quasiparticle, namely the Bogoliubov quasiparticle.  However, the spectrum of excitations includes also vortices, plasmons (phase fluctuations of the order parameter), and the Higgs mode (amplitude fluctuations of the order parameter). The Higgs mode does not contribute to any conventional form of linear response\cite{Pekker2015}. However, we still have to account for the vortices and the plasmon.

At first glance one could say `we wrote down the current operator in terms of electrons (Eq. \ref{barecurrent}), and then solved the problem with this starting point, so what could we possibly have missed?' However, the current operator in a superconductor has an additional piece not accounted for in Eq. \ref{barecurrent}, namely the diamagnetic piece $\vec{j}_d \sim \vec{A}$ (where $\vec{A}$ is the vector potential). In superconductors, the vector potential includes also a `longitudinal' polarization, which is a relic of the plasma oscillation mode in the metal. It is this diamagnetic piece that gives rise to the supercurrent response (i.e. to the $\delta(\omega)$ term in the real part of the conductivity), and it is here that the effects of vortices and plasmons could be lurking. However, it is straightforward to see that this piece (or at least the smooth part of it, which corresponds to plasmons) cannot contribute to the ac conductivity in linear response. In brief, since this term in the current operator is already linear in the vector potential $A$, inserting it into the Kubo formula will give a contribution at order $A^2$, which goes beyond linear response. Such a term can contribute to linear response only through its quantum expectation value $\langle A \rangle$ evaluated in the unperturbed state of the system, and such a quantum expectation value can only contribute to the dc response, not to the ac response. Thus it is straightforward to see that the plasmons do not contribute to the ac conductivity in linear response.

The singular part of the $A$ field (i.e. the vortices) require a little more care. If we are in the Meissner phase, then vortices are absent, and do not contribute to conductivity. This is true even at non-zero temperature, since a single vortex threading across the system costs an energy $\sim L$, where $L$ is system size. In the thermodynamic limit, vortices threading across the system are thus infinitely energetic, and not part of our spectrum of excitations. (Finite size vortex loops will of course arise at non-zero temperature, but these will feel zero net Magnus force in the presence of a supercurrent, and thus will not give rise to any resistance). Thus in the Meissner phase, vortices will not alter the results that we have derived. The real part of the conductivity will contain a delta function at zero frequency (from supercurrent response), plus a finite frequency piece that will vanish as a power law $\omega^{\phi}$, potentially with $\phi>2$ depending on the symmetry class and the temperature. 

The situation is different if we assume we are working with a type II superconductor in an applied magnetic field $H_{c1} < H < H_{c2}$, such that there is a non-zero density of vortices threading the system. In this case, the response will depend on whether the vortices are delocalized (in which case the coupling between vortex and quasiparticle sectors must be fine tuned to zero, otherwise the quasiparticles will delocalize), or whether the vortices are themselves localized. The latter possibility (vortex localization) was discussed in Ref. \onlinecite{extended}. If the vortices are delocalized, then a current applies a Magnus force on vortices, which drives `flux flow' transverse to the current. This flux flow in turn induces an electric field against the current, endowing the system with a non-zero resistance in the low frequency limit \cite{Tinkham2004}. The `super-current response,' which consists of a delta function at zero frequency in the absence of vortices, will be broadened into a Drude peak, and will swamp the signal we have described above. Thus, our results are not expected to apply in the presence of delocalized vortices. If vortices are {\it localized}, then while an infinitesimal AC current will still produce an infinitesimal force on vortices, this force will not drive flux flow. In the absence of flux flow, the system will continue to have zero resistance in the low frequency limit, so the `supercurrent' delta function at zero frequency will not be broadened into a Drude peak. The finite frequency conductivity will therefore continue to be dominated by the quasiparticles, and will take the form calculated above. While the arguments given here assume closed system quantum dynamics, analogous results obtain if we instead assume that vortices are in the vortex glass phase of an open quantum system (see Ref.\onlinecite{FFH}), in which case the vortices are pinned on disorder. A {\it finite} AC current may well drive depinning phenomena, but these will lie beyond linear response. 

Thus, our results are expected to apply both in the Meissner phase, and in the intermediate regime of a type II superconductor if the vortices are localized on disorder. They are not expected to apply in the intermediate phase of a type II superconductor if the vortices are delocalized. Our results are also particular to linear response. The contribution of localized vortices (and potentially plasmons) to {\it non-linear} response AC conductivity is an interesting open problem that we leave to future work. Our results are also particular to the {\it bulk} response of the superconductor. The contribution of boundary currents to the response of a superconductor will be discussed at length elsewhere \cite{edge}. 

Finally as with any localized system, one must consider the effect of rare regions with low disorder, also known as Griffiths regions. It has been argued that in spatial dimensions greater than one, such rare regions destabilize localization on timescales exponentially long in disorder strength \cite{stability}. Such rare regions may thus alter our scaling behavior on frequency scales exponentially small in disorder strength. It has also been argued that rare regions can modify the scaling form of the ac conductivity close to a delocalization transition \cite{gopal}. All considerations of rare region effects are beyond the scope of the present paper, which concentrates on phenomena at typical points in space.  

To conclude, in this paper, we have discussed the low frequency AC conductivity of a many-body localized superconductor. In the Anderson-localized limit, we have argued that the system can display an unusual $\sigma(\omega) \sim \omega^{\phi}$ scaling with $\phi>2$ at zero temperature. This behavior is argued to be robust to interactions at zero temperature. At non-zero temperature and with interactions there is a crossover between a low frequency regime which is of the form discussed in Ref.\onlinecite{gopal}, and a broad intermediate frequency regime with unusual power laws. We have discussed this crossover at length and have shown that it is controlled by the average number of flips contained within a localization volume, a quantity which we denote as $\alpha^{-1}$.

Our arguments are not expected to apply to the intermediate phase of type II superconductors with mobile vortices. However they are expected to apply both to the Meissner phase, and to the intermediate phase when vortices are localized on disorder. In these regimes they provide a characterization of localized superconductors through their ac conductivity, a quantity readily accessible through optical experiments. Our results are also particular to linear response, and to typical points in space. Consideration of effects beyond linear response, and also considerations of rare regions effects and how they modify the behavior discussed herein, would be worthwhile projects for future work. 

\section*{Acknowledgments}

We acknowledge useful conversations with Yang-Zhi Chou, Leo Radzihovsky, Senthil Todadri, Victor Gurarie, and Sarang Gopalakrishnan.  We also thank an anonymous referee for pointing out an error in the first version of this manuscript.  This material is based in part upon work supported by the Air Force Office of Scientific Research under award number FA9550-17-1-0183 (ATS and RMN). MP is supported partially by a Simons investigator award to Leo Radzihovsky, and partially by the NSF Grant 1734006.

\normalem

\end{document}